# OPEN-SOURCE MATLAB-BASED GRAPHICAL USER INTERFACE (GUI) FOR COMPUTER CONTROL OF MICROSCOPES USING μMANAGER

Quang Long Pham,[1] David Chege,[2] Timothy Dijamco,[3] Migle Surblyte,[3] Akshay Naik,[2] Kolawole Campbell,[2] Nhat-Anh-Nguyen Tong,[1] Roman Voronov[1,*]

[1] Otto H. York Department of Chemical and Materials Engineering, Newark College of Engineering, New Jersey Institute of Technology, Newark, NJ 07102, USA

[2] Department of Electrical and Computer Engineering, Newark College of Engineering, New Jersey Institute of Technology, Newark, NJ 07102, USA

[3] Department of Computer Science, Ying Wu College of Computing, New Jersey Institute of Technology, Newark, NJ 07102, USA

[*] Corresponding Author

Email Address: *long.pham@njit.edu* (Q.L. Pham), *chegedm909@gmail.com* (D. Chege), *tmd24@njit.edu* (T. Dijamco), *ms2286@njit.edu* (M. Surblyte), *an367@njit.edu* (A. Naik), *koc3@njit.edu* (K. Campbell), *at562@njit.edu* (N.A.N Tong), *rvoronov@njit.edu* (R. Voronov)

## ABSTRACT

Live time-lapse microscopy is essential for a wide range of biological applications. Software-based automation is the gold standard for the operation of hardware accessories necessary for image acquisition. Given that current software packages are neither affordable nor open to complex structured programming, we have developed a Matlab-based graphical user interface (GUI) that is fundamentally accessible while providing limitless avenues for further customization. The GUI simultaneously communicates with the open-source, cross-functional platform μManager for controlling hardware for time-lapse image acquisition and with other software for image processing. The use of the GUI is demonstrated through an 18-hour cell migration experiment. The results suggest that the GUI can generate high-quality, high-throughput time-lapse images. The core code behind the GUI is open-source so that it can be modified and upgraded. Therefore, it benefits the researchers worldwide by providing them a fundamental yet functional template to add in advanced features specific to their needs, such as additional non-microscopy hardware parts and software packages supported by the Matlab software.

## INTRODUCTION

Live time-lapse microscopy is essential for a wide range of biological applications, including studying tissue engineering cultures as well as monitoring cell signaling and behavior.[1-3] Unlike sacrificial analysis methods, where results for each time point are collected from a *different* sample, time-lapse microscopy provides a *continuous* record of the experiment. This yields more complete information about the biological phenomena at hand. However, given that most biologically-relevant time scales require long-term experimentation, manual time-lapse microscopy becomes too cumbersome for human labor. Consequently, software-based automation offers a solution for overcoming this problem.

However, a big challenge for such software is that biological experiments typically involve multiple hardware components in addition to the microscope itself (e.g., motorized stage, filter turret, fluorescence light source, laser, and digital camera.) These components are often made by different companies, each of which provides proprietary drivers and software specific to the devices that they sell. Hence, the automation software must act as an "umbrella" that supports the individual drivers that come with the hardware, and should be able to synchronize all these parts .

Several such commercial software packages exist, such as MetaMorph7 (Molecular Devices, San Jose, CA), Element 4 (Nikon, Tokyo, Japan), iQ 2.6 (Andor, Belfast, UK), and cellSens (Olympus, Japan). Although these can drive a large number of *microscopy* devices and provide some basic image processing capabilities, they also have significant disadvantages: (1) prohibitive cost, (2) lack of complex programming capabilities, such as hardware automation, image analysis, and mathematical modeling, and (3) inability to communicate with *non-microscopy* hardware, such as solenoid valves, syringe pumps, culturing chambers, and other electronic devices that are commonly used in biological experiments (e.g., microfluidics).

In addition to these commercial solutions, µManager (MM) is a free option that supports a wide range of *microscopy* hardware devices. Like the commercial analogues, it comes with its own graphical user interface (GUI) that provides basic experiment setup and data post-processing capabilities.[4] Consequently, it has become a popular tool for microscopy imaging.[5, 6] However, the MM GUI is limited to basic experiment customization/automation functionality, and only a few additional features are provided via external plug-ins. It does not allow for complex structured programming, or non-microscopy functions that could be useful, such as image processing, mathematical analysis and the ability to communicate with *non-microscopy* hardware.[7] For these purposes, application programming interfaces (API)[8] are provided that allow the user to interface with more advanced programming environments, such as Java, C++, and Matlab®.[9]

Among these, Matlab enables a much wider range of experiment customization via toolboxes like Computer Vision and Image Processing, Supercomputing, and Machine Learning. Another advantage of Matlab is that it offers support for a wide range of *non-microscopy* hardware.[10] Matlab offers support for ARM, Arduino, Altera, National Instruments, Raspberry Pi, Xilinx, Android, STMicroelectronics, and other devices that are frequently integrated into biological experiments. Finally, despite being a commercial package, Matlab is available to most researchers through their universities. For these reasons, Matlab is the natural choice for implementing the advanced automation of microscopy experiments via MM.

Despite the benefits of this method, developing custom Matlab code which sends commands to the microscopy hardware via MM is an impractical undertaking for most labs. Development can be time intensive and challenging because there are only a few examples (most of which are very simple) available in the online Matlab documentation. Furthermore, most of the basic functionality around the microscope (e.g., capturing and stitching images, moving the stage, changing objectives, compensating for vibrations and artifacts, etc.) should be the same for different labs, rendering the work of creating the code redundant. Finally, even if one is successful in writing such Matlab-MM automation code, it would likely be difficult for other lab members who lack an advanced programming background to use it. Hence, a Matlab-MM GUI would be ideal for simplifying the end-user experience, while also retaining the flexibility of utilizing advanced Matlab features for experiment customization and automation. Yet, adding a GUI on top of Matlab-MM is even more labor-intensive and redundant. Hence, the microscopy community would benefit from a freely-available Matlab-MM GUI that contains all of the basic features expected of microscopy software. This would save researchers time by providing them with a substantial starting point that they can then build upon in order to meet their specific needs.

To that end, this manuscript offers such a Matlab-MM GUI to the public as an open-source code. The GUI contains a wide variety of fundamental microscopy control and image acquisition features, as well as some additional post-processing functions. Some of the latter include stitching multiple tiles into a single panorama and light intensity flattening. Additionally, the GUI offers an implementation of a novel tile-based video stabilization algorithm. Although the code was developed using the Olympus IX83 platform, in principle it should be compatible with any other hardware brands as long as they are supported by MM. This code will benefit researchers by providing them with a template for adding their own features into it. Some examples of the possible advanced customizations using Matlab include enhancing the experiment via support for additional electronic devices and/or with feedback loops based on real time image interpretation via computer vision; increasing computational processing power by off-loading data analysis to supercomputers and then retrieving the results automatically; and supplementing experimental analysis with COMSOL® Multiphysics simulations using the Matlab Livelink® interface. Not only would these enhancements save time for others, but they would also open up limitless new possibilities in terms of experiment design.

**METHODS**

*Hardware and Software*
Our GUI was developed on hardware commonly used by biological research laboratories. As noted earlier, our code is compatible with any hardware supported by MM. The hardware mentioned in this text serves were chosen as an example to demonstrate our software's capability. The individual components used in the setup include an inverted two-deck microscopy system (IX83, Olympus, Japan), an XY motorized linear-encoded translational stage (96S106-O3-LE2, Ludl, NY), a digital complementary metal-oxide semiconductor (CMOS) camera (Orca Flash V4.0, Hamamatsu, Japan), and a custom-built workstation with 64-bit operating system Microsoft Window 8.1, an Intel Xeon E5-2650 Processor (2x 2.00 GHz), 16 GB of DDR3 RAM, and a 512 GB SATA 2.5" solid state hard drive. The Ludl stage is driven by a motor drive controller (MAC 6000, Ludl) which is connected to the computer via universal serial bus 2 (USB2) interface. The camera is connected to the computer via USB3. The microscope itself is a fully automated platform that is comprised of several modules that can communicate with a computer via a firewire interface. In particular, it includes: a motorized long working distance (LWD) condenser (IX3-LWUCDA, Olympus) which can switch between phase contrast and bright field, but can also act as a shutter for the camera; a motorized/coded fluorescent mirror turret (IX3-RFACA, Olympus); the microscope body (IX83P2ZF), which contains a motorized nosepiece for switching between objective lens as well as moving the objectives in Z direction for sample focusing; an LED lamp house (IX3-LHLEDC) whose illumination power can be controlled externally; and an autofocus module called the Z-drift compensator (IX3-ZDC), which can search for the shift in the imaging plane and compensate for it during the course of the experiment. The microscope uses objective lenses of various magnifications for different purposes. There are 1.25X objective lenses (PLAPO1.25X, NA 0.04, Olympus) for fast acquisition of a large "overview" area for region of interest (ROI) selection, 10X phase contrast objectives (UPLFLN10X PH1,

NA 0.30, Olympus) for acquiring the phase contrast image, and 60X objectives (LUCPLFLN60XPH, LWD NA 0.70, Olympus) which work with the ZDC module for autofocus.

The latest version of μManager (Version 2.0 Beta) does not support the Olympus IX83 microscope, which is a popular brand and has been widely used in biological research. In order to make our code compatible with a wider range of microscopes, we built it using the older version 1.4 of μManager. The GUI was built using Matlab R2016b, but a newer version of Matlab can still be used with minimal modification of the code. The GUI was produced using a Matlab Tool called the GUI Development Environment (GUIDE). The GUI elements (i.e. buttons, check boxes, message boxes, dialog boxes, and drop-down list) are designed and organized using the GUIDE layout. The elements' functions are defined using the command window of the Matlab GUI. Postprocessing is done with ImageJ version 1.5n, a free open source image processing tool.[11]

*Microfluidic Device Fabrication*

The microfluidic device was fabricated from a silicone elastomer poly(dimethyl siloxane) (PDMS, Sylgard 184, Dow Corning, MI) using soft lithography techniques. The master mold for the device was generated using a negative photoresist (SU-8 2015, MicroChem, MA). First, the microscale pattern was sketched using AutoCAD (Autodesk, Mill Valley, CA) and printed at 16,525 dpi on a transparency to generate a high-resolution photomask. The SU-8 was spin coated, exposed to UV light, and developed on a 4-inch silicon wafer to generate micron-sized channels. Then, PDMS with a base-to-agent ratio of 10:1 was poured over the mold and cured at 65 °C overnight. The cured PDMS was removed from the master mold and treated with air plasma prior to being bonded to a glass substrate. Right after bonding, the channels were coated with poly-D-lysine to enhance cell attachment.

*Cell Preparation and Culturing*

Mouse embryo NIH/3T3 (ATCC® CRL-1658TM) fibroblasts were purchased from ATCC (Manassas, VA). Prior to being transferred to the microfluidic device for the migration experiments, the cells were incubated in culture media inside of T75 flasks. The flasks were kept at 37 °C and in a humidified atmosphere of 5% $CO_2$ in air. The culture media was changed every two days to ensure normal cell growth. Prior to the migration experiments, the cells were trypsinized from the T75 flasks and loaded into the devices, with a seeding density of about 50,000 cells $cm^{-2}$. The devices were incubated at 37 °C under 5% $CO_2$ for 6 h to allow cell attachment. Then, the cells were cultured in serum-starved media (MEM supplemented with 1% penicillin-streptomycin) for 6 hours.

*Cell Migration Experiments and Image Acquisition*

At the start of the experiment, the cell culture media in the chip was replaced with $CO_2$-independent basal media buffered by HEPES. 20 μL of basal media supplemented with 50 ng $mL^{-1}$ PDGF-BB was then added into the central reservoir of each device. The devices are placed inside a condition chamber maintained at 37 °C mounted on top of the motorized stage of the microscope. Time-lapse phase-contrast imaging of the fibroblast migration was performed using the 10X phase-contrast objective. Images were automatically captured at 10 min intervals for the duration of 18 hours. For each device at each time step, 25 or 36 tile images were acquired at various locations, with autofocus enabled. Images are stitched and stabilized using the GUI.

**DESCRIPTION OF TILE-BASED IMAGE STABILIZATION ALGORITHM**

Time-lapse microscopic imaging is prone to stage drift, which can impact the final quality of an acquisition. The situation is further complicated by the fact that a single frame could be composed of many image tiles, which together may be too large to stabilize at once. In order to improve the quality of images acquired with this method, there is a need for a memory efficient method to correct for stage drift after imaging.

Our GUI features a novel image stabilization algorithm which can correct the effects of stage drift while lowering memory footprint. Instead of correcting for drift by comparing the entire frame from one timestep to the next, our algorithm analyzes image tiles, calculates their correlation, and finds the average transformation for the most correlated tiles. This allows us to create high quality image stabilization while lowering the memory footprint.

*Estimating individual tile drift*

The stabilization algorithm first estimates the drift of each individual image tile between each time step. To do this, an affine transformation matrix is calculated for each tile and time step, beginning on the second time step. Each transformation matrix serves to estimate how that image tile has drifted between the previous time step and the current time step. We begin calculating the matrices on the second time step because images from the first time step cannot be compared to a previous image.

To calculate the transformation matrix for any particular image tile $I$ at time step $t$, $I_t$ and $I_{t-1}$ are passed into the function compareImagesForStabilization. This function implements the Speeded Up Robust Features (SURF) algorithm, which uses a Fast-Hessian detector to locate features of interest within the image tiles. The strongest 100 of these SURF features in each image are extracted into feature vectors, and the features that can be matched between the two images are found. Any matched points between which the distance is more than 50 are assumed to be errors and are eliminated. The remaining points are passed into the function estimateGeometricTransform, which calculates the affine transformation matrix for that pair using the M-estimator SAmple Consensus algorithm. The algorithm only uses the coefficieints for horizontal and vertical movement, so all extraneous coefficients within the generated transfomation matrix are disabled by setting their value equal to 1. As they are created, each transformation matrix is stored in a four-dimensional matrix which will eventually hold all of the transformation matrices.

Since we only need the translation coefficients in the transformation, the program reduces the four-dimensional matrix to a three-dimensional matrix that stores only these coefficients.

*Calculating correlations*

After all of the transformation matrices are calculated, the program constructs a correlation coefficient matrix, denoted **C**, correlating the drift in the x direction among all of the image tiles. Given $n$ as the total number of image tiles per timestep, **C** of size $n \times n$. The program iterates over each pair of tiles and constructs **C** such that

$$\mathbf{C} = \begin{pmatrix} c_{1,1} & c_{1,2} & c_{1,3} & \ldots & c_{1,m} \\ c_{2,1} & c_{2,2} & c_{2,3} & \ldots & c_{2,m} \\ c_{3,1} & c_{3,2} & c_{3,3} & \ldots & c_{3,m} \\ \ldots & \ldots & \ldots & \ldots & \ldots \\ c_{n,1} & c_{n,2} & c_{n,3} & \ldots & c_{n,m} \end{pmatrix} \qquad (1)$$

Let $A$ be the dx coefficient (translation in the x direction) for the $n$th tile, and let $B$ be the dx coefficient for the $m$th tile. Each element $c_{n,m}$ of the matrix **C** can be defined as

$$c_{n,m} = \rho(A,B) = \frac{\text{cov}(A,B)}{\sigma_A \sigma_B}, \qquad (2)$$

where $\rho(A,B)$ is the Pearson correlation coefficient of the pair of dx coefficients. This calculation is performed using the Matlab function corrcoef.

*Calculating the desired translation of each whole stitched image*

In order to locate the largest set of tiles with correlated dx coefficients, the program repeatedly performs a Dulmage-Mendelsohn decomposition on the portion of the correlation matrix which is above a given threshold, using the Matlab function dmperm. On each iteration, the program relaxes this threshold for tiles being "closely correlated". The loop ends once the program has found a correlated group that contains at least three tiles. This process detects groups of tiles which have closely correlated dx coefficients. A new coefficient matrix is created containing only the tiles that are in this group.

For each time step, the dx and dy coefficients among all the closely correlated tiles are averaged. A matrix is created which stores the average dx and dy coefficients for each time step. This newly constructed matrix, denoted **D**, is constructed such that

$$\mathbf{D} = \begin{pmatrix} D_{1,1} & D_{1,2} \\ D_{2,1} & D_{2,2} \\ \ldots & \ldots \\ D_{t,1} & D_{t,2} \end{pmatrix}, \qquad (3)$$

where $D_{t,1}$ is the average dx coefficient and $D_{t,2}$ is the average dy coefficient for timestep $t$. Let $dx_t$ and $dy_t$ be the dx and dy coefficients at timestep $t$, respectively. An element of **D** is defined as

$$D_{t,1} = \frac{1}{t}\sum_{i=1}^{t} dx_i \tag{4}$$

and

$$D_{t,2} = \frac{1}{t}\sum_{i=1}^{t} dy_i \tag{5}$$

*Transforming each stitched image*

The program iterates through the time steps, beginning on the second time step because there is no reference point against which to stabilize the first timestep. A 3x3 affine transformation matrix, denoted by **E**, is created for each time step using the translation coefficient values that were calculated in **D**. This matrix is then multiplied with a cumulative transformation matrix, which keeps track of how much correction is needed relative to the first time step. The original image is loaded and the appropriate transformations are applied using Matlab's imwarp function. The stabilized image is saved as a TIFF file.

**DESCRIPTION OF THE GUI AND INSTRUCTIONS FOR ITS USE**

The GUI code can be readily downloaded from: https://git.njit.edu/rvoronov/ROBOSCOP.git, while MM is available at: https://micro-manager.org/wiki/Download%20Micro-Manager_Latest%20Release  For the latter we recommend version 1.4.21, as it is the latest to support the Olympus IX83 microscope.

*Setting up the μManager-Matlab Environment*
The following instruction show how set up Matlab interfacing with μManager for Windows computers (adapted from μManager web page) [12, 13]

1. Install μManager (version 1.4) to "C:\Micro-Manager-1.4". From Windows, open the Command Prompt and type in "cd C:\Micro-Manager-1.4".  Then, use the following command to save a list of all *.jar files in the folder: *dir /s /b /o:gn *jar > jarlist.txt*. This will create a new file called jarlist.txt in the current folder.

2. Open Matlab and run : *edit([prefdir '/javaclasspath.txt']);*. This opens a .txt file within the Matlab interface. Copy all of the filenames from jarlist.txt and paste them into the .txt file opened by Matlab. Save and close the file.

3. In Matlab, type: *edit([prefdir '/librarypath.txt']);*. Copy the μManager installation path (i.e. "C:\Micro-Manager-1.4") to this .txt file. Save and close the file.

4. Restart Matlab. μManager can now be started through Matlab using the following two commands (already integrated into our code):

```
import org.micromanager.MMStudio;
gui = MMStudio(false);
```

5. Set up the MM hardware configuration wizard (the GUI code obtains the objective labels from the MM hardware configuration profile *.cfg).

6. Finally, run the OLYMPUS_MICROMANAGER_GUI.m using Matlab in order to initiate the Matlab-MM GUI.

The structure of the GUI is shown in Figure 1. Different elements of the GUI and their functions are described below:

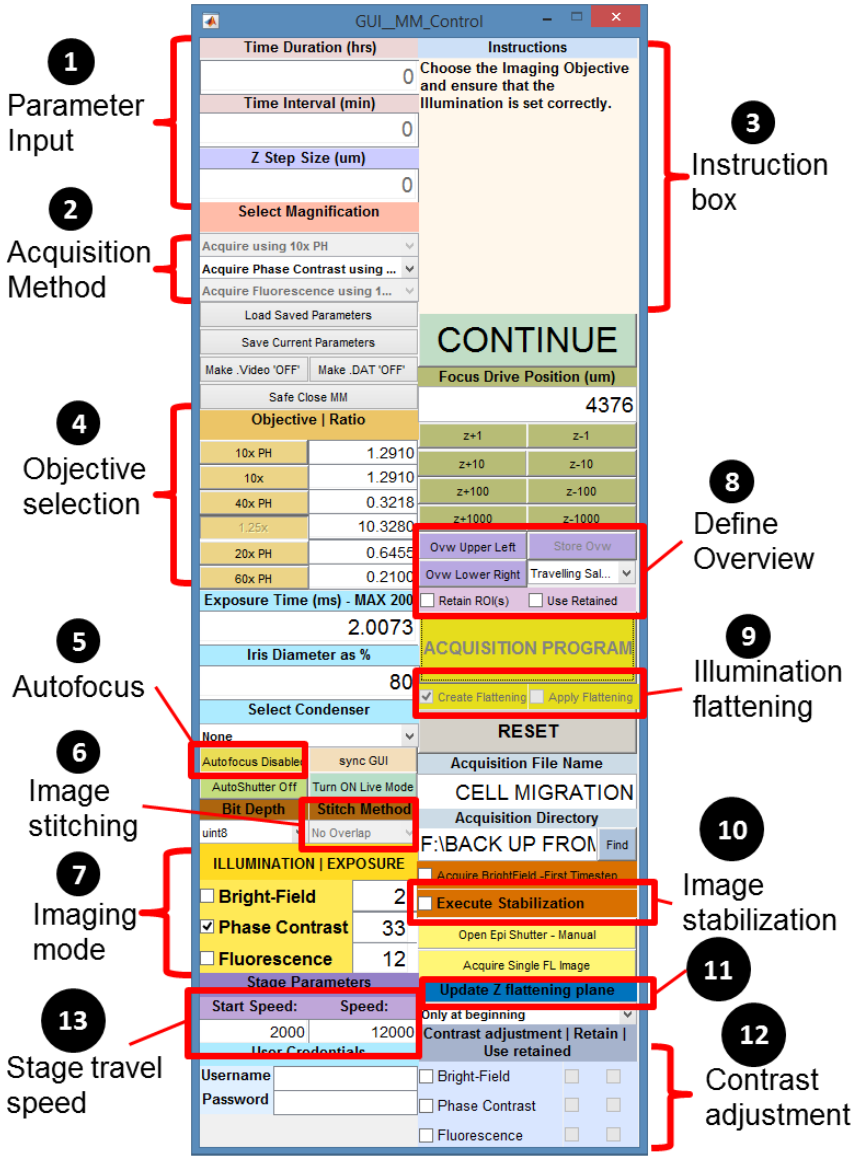

**Figure 1** Structure of the Matlab-based Graphic User Interface with different options and functions

1. PARAMETER INPUT: Dialog boxes that allow the user to input parameters such as the duration of imaging, time interval, and Z step (the distance interval the objective lens moves in the Z direction between a defined travel range). Usually, for 2D imaging in which the objective lens is fixed in the Z direction, the Z step is set to 0. In the case of imaging at different Z planes (i.e. for 3D construction), the Z step needs to be defined by the user. Z step is set in µm.
2. ACQUISITION METHOD: By default, the image acquisition is performed using a 10X objective for all of the imaging modes. If the user wants to image to a different magnification, they can select it from the drop down list.
3. INSTRUCTION BOX: A static box displaying instructions to guide the user through each step of the set up process.
4. OBJECTIVE SELECTION: Buttons allow the user to manually switch between objective lenses without actually touching the microscope hardware. Each number in the "Ratio" column indicates the pixel size (in μm) of the image acquired using the corresponding objective lens. (*Note:* the micron-to-pixel ratios are either calibrated or estimated in MM, and subsequently pulled from the hardware configuration *.cfg file by the GUI)
5. AUTOFOCUS (OPTIONAL): A button that, once enabled, can send Matlab commands to the Z drift compensating module via µManager. The Autofocus function needs to be activated by checking "Focus Limit Settings" on the microscope control pad (Figure 2A). This is indicated by the yellow line with a lock symbol on the bottom right corner of the acquisition window of the control pad (Figure 2B). Even though this setting is specific to the Olympus IX83 microscope used in our setup, the setting procedure for other microscope brands should be analogous. Note: the GUI assumes that the *last* position in the motorized objective nosepiece is an autofocus-compatible objective (consult the autofocus instruction manual to check compatibility).

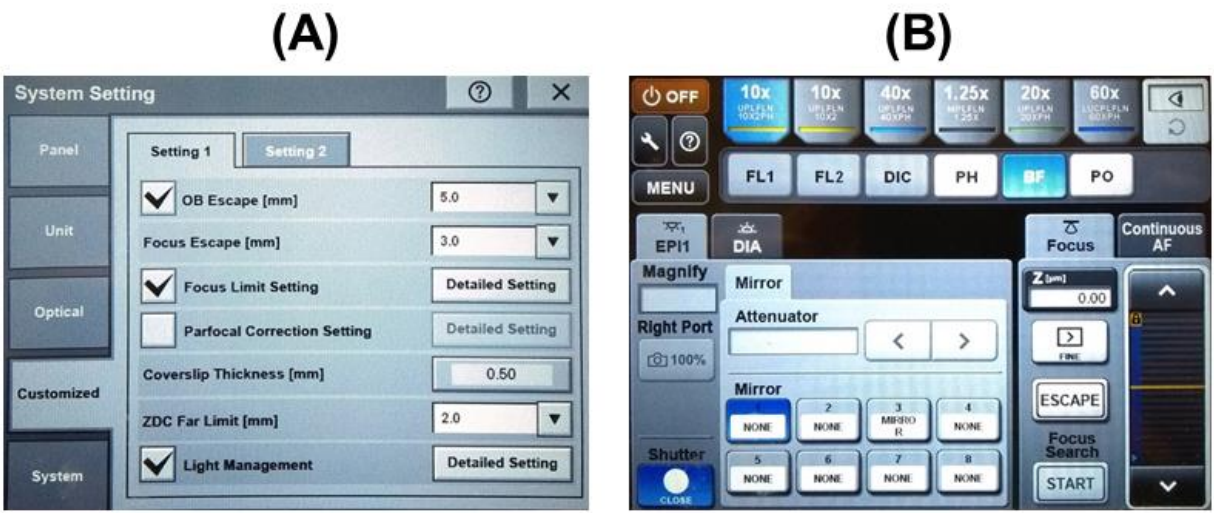

**Figure 2** Focus Settings on the touch pad of the IX83 microscope. (A) In order to activate the Z drift compensating IX3-ZDC module, Focus Limit Setting needs to be checked. (B) The main control panel showing a yellow line at a certain Z position. This indicates the ZDC has been activated and is ready to be controlled by the GUI.

6. IMAGE STITCHING: A drop down list allows users to select different stitching methods. There are non-overlap and overlap stitching options. In the former method, the tiles have no overlapping in the field of view (FOV). Hence, the stitching is done by merely positioning the tiles next to each other in the right order to form the panorama image. The latter, on the other hand, overlaps the tiles' FOV in order to produce seamless results by having the overlapping regions (20% overlap is chosen as default) serve as "registration marks" for the XY alignment of one tile onto another. Two options are available for the overlap stitching: "Grid: using BF" (which stands for bright field) and "Grid: using PC" (which stands for phase contrast). While the BF option relies on bright field imaging for the point feature matching, the latter is based on images collected from phase contrast microscopy. There is no option for stitching based on fluorescence (FL) tiles, since FL generally does not give sufficient information for registering all of the tiles. However, either BF or PC can be used to aid the stitching of fluorescence images. (*Note:* the stitching algorithim does not compensate for tile $\theta$-rotation relative to each other. Hence, the orientation of the camera in the microscope's light port needs to be tuned manually by trial-and-error, using a sample with a distinct pattern across all tiles).
7. IMAGING MODE: Check boxes that allow users to select different imaging modalities (BF/PC/FL), and a dialog box that allows them to input the desired exposure time for each.
8. DEFINE OVERVIEW: A total imaging area is created using a low-magnification objective (1.25X by default, but could be chosen to be anys). Users need to define the upper left and lower right corner of an overview image by manualy moving the stage. A dropdown list allows users to select different travelling modes to optimize the travel of the XY stage (i.e. obtain the same number of images using the shortest travel path length or a user-specified one). Users can also either select the "Retain ROI(s)" option to restore the selected coordinates for future use, or "Use Retained" to skip the ROI selection by using coordinates restored from a prior aquisition.
9. ILLUMINATION FLATTENING: Users can select: (1) "Create Flattening", to create a reference "background" image for correcting Vignette illumination artifacts; or (2) "Apply Flattening", to subtract the background from current image using the reference created in (1).
10. IMAGE STABILIZATION: Once "Execute Stabilization" is checked, a Matlab function will be called to initiate the stabilization algorithm, previously outlined in the DESCRIPTION OF TILE-BASED IMAGE STABILIZATION ALGORITHM section. This option will generate a series of stabilized time-step images, which will be saved in "Stabilized" folder, located in the same folder that contains the stiched images.

11. UPDATE Z FLATTENING PLANE: The user can specify how frequently the autofocus takes place during the acquisition by selecting from a dropdown list the "x" number of time steps between two consecutive autofocus events. The hardware autofocus uses a laser beam to measure the distance from the objective to the sample, and then updates the objective height accordingly to keep the sample in focus. In our code, the measurement takes place at the four corners of the user-defined "overview" region (see Step 8 above). To that end, the frequency with which these measurements are updated throughout the aquisition is controlled by the user here. For example, if "only at beginning" is selected, the autofocus only happen before the 1$^{st}$ time step, while "every 5 time steps" means the autofocus will happen every 5 time steps during the whole imaging sequence. Once the autofocus measurements are obtained, a 3D plane is fitted to these points, and is subsequently used to interpolate the appropriate Z height of the objective at every individual tile (see section 2.12 for additional information). The same interpolation plane is used until it is updated with a subsequent autofocus measurement. If more than 3 autofocus measurements fail for any reason, the code defaults back to the interpolation plane from the most recent successful measurement.
12. CONSTRAST ADJUSTMENT: This function allows the user to adjust the contrast of each ROI and each imaging mode (i.e. BF, PC, and FL) before a time-lapse acquisition is run. This function is needed to prevent the final image from being too bright or too dark, according to the user's preference.
13. STAGE TRAVEL SPEED: Here the user can manually input the travelling speed of the motorized stage. By increasing the travel speed, the user can achieve faster image acquisition. However, higher travel speed may lead to some issues with the imaging, such as blurring of the images. Moreover, the inertia force that may disturb the specimen (e.g. causing waves of liquid), especially at a high frequency. Hence the stage's speed should be chosen while keeping in mind the camera's exposure time as well as the overall experimental design.

The GUI also offers some basic control of the microscope without the user needing to touch the microscope's hardware. These controls include moving the objective in the Z direction, switching between objective lenses, swapping condenser optics, and changing the illumination power and the opening size of the iris diaphragm.

**HARDWARE SYNCHRONIZATION AND SOFTWARE COMMUNICATION**

The synchronization between software and hardware is illustrated in Figure 3. The Matlab-based GUI plays a central role in orchestrating all of these components by simultaneously communicating with μManager for hardware control while interfacing with ImageJ for image post-processing. The Matlab-MM interfacing allows the images to be acquired at different XY positions, Z planes, time intervals, and microscopy modes. The communications between μManager and the XY stage include toggling XY translation, real-time updating of sample position, and changing travel speed and acceleration. The GUI interfaces with the camera in capturing images. In the case of bright field and phase contrast imaging, the image is captured using the camera's electronic shutter mode only to optimize the imaging speed, while in the case of fluorescence microscopy, the physical shutter of the microscope is used during the acquisition (i.e. between two consecutive stage's travel or between time steps) to avoid photobleaching of the sample. The GUI-microscope interaction includes controlling illumination, changing the Z position of the objectives, rotating the motorized objective nosepiece and the filter cube turret, swapping optics in the condenser turret, and opening/closing the epifluorescence shutter.

There are some basic image processing options that users can activate using the GUI. The processing requires communication between the GUI and other software. Image stitching is done using the Collection Stitching Plugin of ImageJ, which is launched using Matlab commands. Image stabilization is done via the Computer Vision System Toolbox in Matlab. Other functions such as rotation and brightness/contrast adjustment, which are created by the Matlab core commands, are also provided by the GUI .

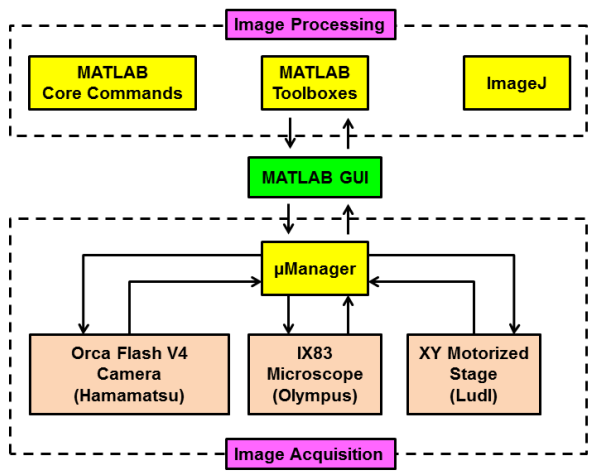

**Figure 3** Block diagram showing communication between software (MATLAB, μManager, and ImageJ) and hardware (automated microscope, motorized stage, and digital camera)

**OPERATIONAL INSTRUCTIONS**

**1. Collect "Background" for Intensity Flattening (Optional)**
Microscope raw images usually come with a vignetted features, which leads to detrimental artifacts on a stitched image. Therefore, a flat-field correction is necessary to remove the vignettes from the image tiles' background prior to stitching.[14] The flattened background for illumination is created using a reference sample. The reference sample can be a clean glass slide in the case of bright field and phase contrast microscopy or a cover glass coated with a layer of fluorescein in the case of fluorescent microscopy. The flattening data is specific to the magnification and illuminiation parameters. Therefore, the objective magnification and its Z position, light intensity, exposure time, condenser iris opening, and microscopy mode (i.e. phase contrast, bright field, or fluoresecence) for creating flattening data must be the same as the one used in image acquisition. The below procedure describes a stepwise process of acquiring flattening data, which is normally done once, unless the imaging parameter is changed:

1.1. In the Select Magnification section, select the right objective from the drop-down list.

1.2. Locate the folder where the raw images of the reference sample will be saved.

1.3. In the GUI, check the "Create Flattening" box, then press "CONTINUE" to go to the next step. A live view window will appear and the instruction box will display a guideline to select the upper left and lower right corner for an overview. By default, the microscope will change to lower magnification (1.25 X objective lens) to gain faster acquisition.

1.4. Move the stage to the upper left and lower right corners of the sample and register their coordinates by hitting the "Ovw Upper Left" and "Ovw Lower Right" buttons, respectively, followed by pressing the "Store Ovw" button. Once the "CONTINUE" button is pressed, the microscope will automatically start acquiring images to build an overview of the selected area.

1.5. Once the overview window pops up, select the ROI(s) by drawing rectangular box(es) around the region where the background data is to be collected (Figure 4). Multiple ROIs may not be necessary in this case; however, the flattening can be improved when more reference images are acquired. After the ROIs are selected, hit "Crop All Selected Regions".

1.6. The microscope will automatically switch to the imaging objective lens (i.e. 10X objective) selected earlier by the user. The sample is brought into focus by adjusting the Z position of the objective. After the minimum and maximum Z is defined, the acquisition begins. The "average" images are calculated from ROIs and stored in the subfolder "resume", which is located in the acquisition folder. These images will be used whenever the "apply flattening" option is selected prior to acquisition.

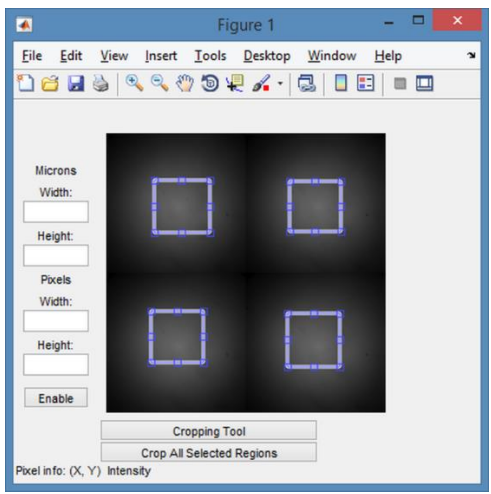

**Figure 4**. Selecting region of interests (ROIs) from an overview. In this example, the overview was stitched from 4 image tiles captured using a 1.25X objective. The artifact in the overview is specific to 1.25 X objective. Hence, it does not affect the flattening data collected by other objectives such as the 10X lens.

**2. Acquire High-Throughput Time-Lapse Images**

The stepwise procedure below demonstrates the use of the GUI in generating time-lapse images of a large sample in a high-throughput manner. Note that all of the defined parameters are specific to this example and are subject to change upon the user's needs.

2.1. Press "ACQUISITION PROGRAM" to initiate the first step of the acquisition.

2.2. Enter an acquisition time of 18 hours, time step of 10 min, and Z step of 0 µm into the "Time Duration (hrs)", "Time Interval", and "Z step size (um)" dialog boxes, respectively.

2.3. In the "Select Magnification" section, select the acquisition method and the magnification from the dropdown list. Select "Acquire using 10X PH" and "Acquire Phase Contrast using 10X objective". These options indicate that the microscope should use the 10X phase objective and condenser lenses during the acquisition.

2.4. Press the "Autofocus Disabled" button to enable the Autofocus module. Once the button is clicked, its label is changed to "Autofocus Enabled".

2.5. Select image quality in under the Bit Depth section and select the stich method (i.e. "Grid: using BF"). The user can select either 8-bit (~ 2MB tile size) or 16-bit (~ 8 MB tile size). The latter comes with 4 times higher file size per image than the former one. However, its disadvantage is that it will quickly consume storage space.

2.6. In the "Illumination | Exposure" area, check the phase contrast and an exposure time of 33 milliseconds.

2.7. Select the stage travelling mode. There are two modes available: "User-defined" and "Traveling Salesman". The former takes into account the order in which the user selects the regions of interest (ROIs) (mentioned in step 2.11 below), and allows the user to manually define the stage traveling route. The "Traveling Saleman" mode, will trigger an algorithm that defines the shortest route for the stage travel (despite of the order in which the user selected the ROIs).[15] This is the most commonly used traveling mode.

2.8. Check "Apply Flattening" to use the flattening data created previously. At the end of each time step, the flattening data is used to generate uniform illumination of the tiles before they are stitched.

2.9. Define the image name in the "Acquisition File Name" box. Specify the location of the image folder in the "Acquisition Directory"

2.10. In the "Update Z flattening plane" section, select "every 5 time steps" to ask the microscope to run the autofocus event every 5 time steps to compensate to any movement/drifting of the sample.

After all the parameters/options are set, hit the CONTINUE button to proceed to the next step. The microscope will automatically switch to the low-magnification 1.25 X objective and µManager's live view window will be turned on, showing the camera view of the sample. Then continue to the following steps:

2.11. Define an overview of the imaging area by moving the sample to the top left position, hitting the "Ovw Upper Left" button, moving the sample to the bottom right position, and hitting "Ovw Lower Right" button. After that, press "Store Ovw" and "CONTINUE" to complete this process. The microscope will start to acquire an overview image of the sample using the low-magnification objective. After the acquisition is finished, all of the image tiles will be stitched to generate a full overview of the selected area (Figure 5A). Define the ROI by drawing rectangles around the interested regions (Figure 5B). Press "Crop All Selected Regions" once all the ROIs are selected to proceed to the next step.

The microscope will switch to the 60X objective lens (LUCPLFLN60x, 0.7 NA, 1.5-2.2 mm adjustable working distance, Olympus) to start searching for the reference focal point (i.e. the top surface of the coverslip where sample is present). Once the focal point is successfully located, the microscope will automatically switch to the acquisition objective lens (i.e. 10X Phase objective).

2.12. Define the maximum and minimum Z positions of the 10X objective. In the case of 2D imaging at a single focal plane, the max and min positions have the same Z value. Thus, for 2D imaging the user just needs to focus on the sample by moving the objective lens in the Z direction to attain the focused image shown on the live view window. However, if Z range needs to be specified, as is the case with 3D imaging, move the objective to the lowest Z value to register the minimum Z and to the highest Z value to register the maximum Z. (*Note: Do not manually move the sample in the XY direction after the autofocus. Otherwise, it will cause a miscalculation of the drift compensation*). After both the max and min Z positions have been specified, the Z positions of four reference points, which are the corners of the overview selected earlier, are determined. This step is necessary to compensate for the initial tilting of the sample. The Z position of individual image tiles can thus be determined by linear interpolation based on their relative position from the reference points. Once all the references have been recorded, a 3D graph illustrating a fitted plane will appear. After that, the acquisition begin. (*Note: these fitting graphs can be found in the subfolder "ZFlattening" located inside the acquisition folder, while detailed coordinates of the 4 reference points can be found in the "AcquisitionLog.txt" file*).

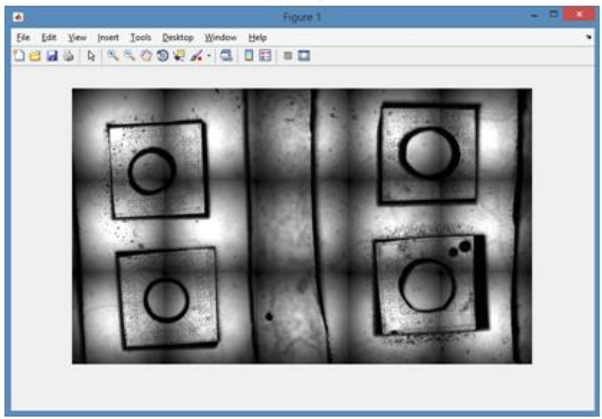 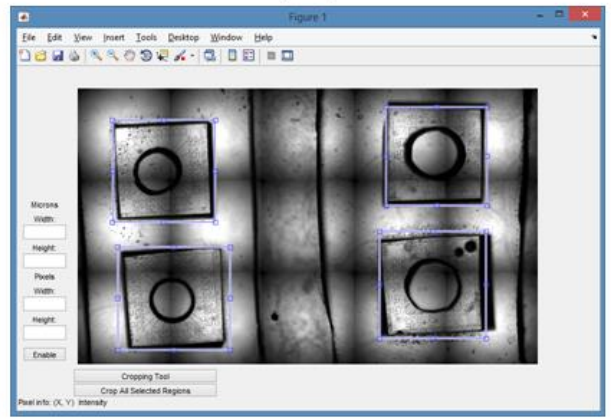

**Figure 5.** (A) Overview of an imaging area acquired using 1.25X objective and stitched using ImageJ plugin. The "upper left" and "lower right" corners of the rectangular area are defined by the user using the GUI. (B) Regions of interest defined by the user. These selected areas will then be "cropped" from the overview, meaning that imaging only happens inside the crops while ignoring the rest of the overview.

## APPLICATION

**Demonstration of the use of the GUI through a cell migration experiment**

In the following section, cell migration (a typical time-lapse imaging experiment,[16-18]) is used to demonstrate the use of the GUI. For this experiment, the chemotaxis of fibroblasts inside a series of microfluidic maze channels is monitored. In order to recruit and guide the cell migration, a concentration gradient of platelet-derived growth factor-BB (PDGF-BB) was generated inside the channels using a flow-free gradient generator. Each microfluidic device contains multiple channels and several devices are placed inside a condition chamber (Figure 6). This allows a high-throughput manner of study. Each selected ROI (step 2.11 above) corresponds to a maze device which has dimensions of 5x5 mm. At 10X magnification, the FOV of the camera is only 1.36x1.36 mm. Therefore, it needs 25 (for "No overlap" stitching type) and 36 (for BF or PC stitching type) tiles to build a panorama of an ROI.

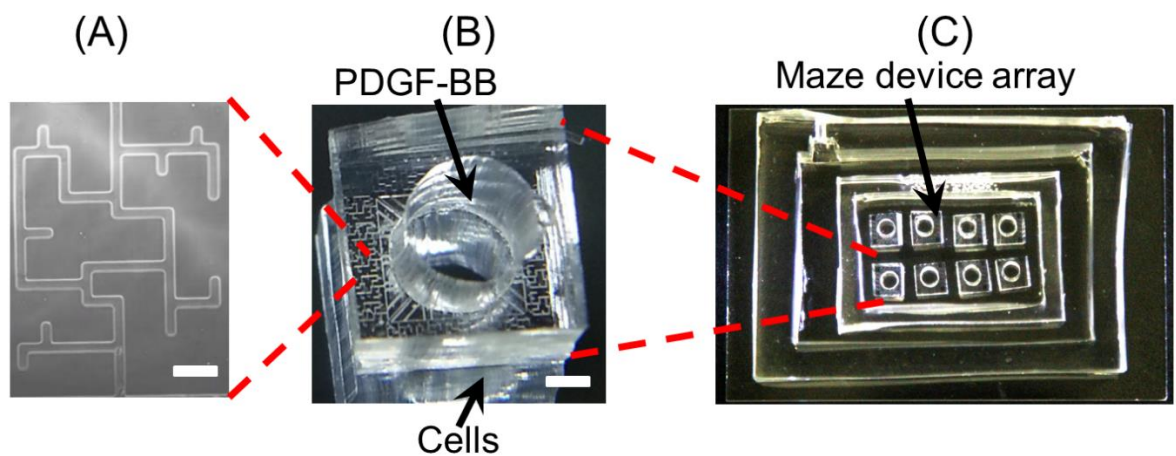

**Figure 6**. Microfluidics maze device used in the experiment. (A) Micrograph of a single maze channels with 24 μm width. Scale bar is 100 μm. (B) Photograph of a single microfluidic device made from PDMS. The central hole is filled with chemoattractant and cells are seeded to the outside. Scale bar is 1 mm. (C) Photograph of a chip which contains an array of maze devices.

The ability of the software to stitch raw tiles to produce a seamless panorama of the ROI is demonstrated. Different stitching methods are compared (Figure 7). The acquisition with "No Overlap" option does not generate tiles with overlapping pixels for registration. Therefore, misalignment occurs in the XY directions when these tiles are stitched (Figure 7, top pane). The "Grid: using BF" and "Grid: using PC" options, on the other hand, yield images with about 20% overlaps, thus being able to create seamless bright field and phase contrast panoramas, respectively (Figure 7, middle and bottom panes).

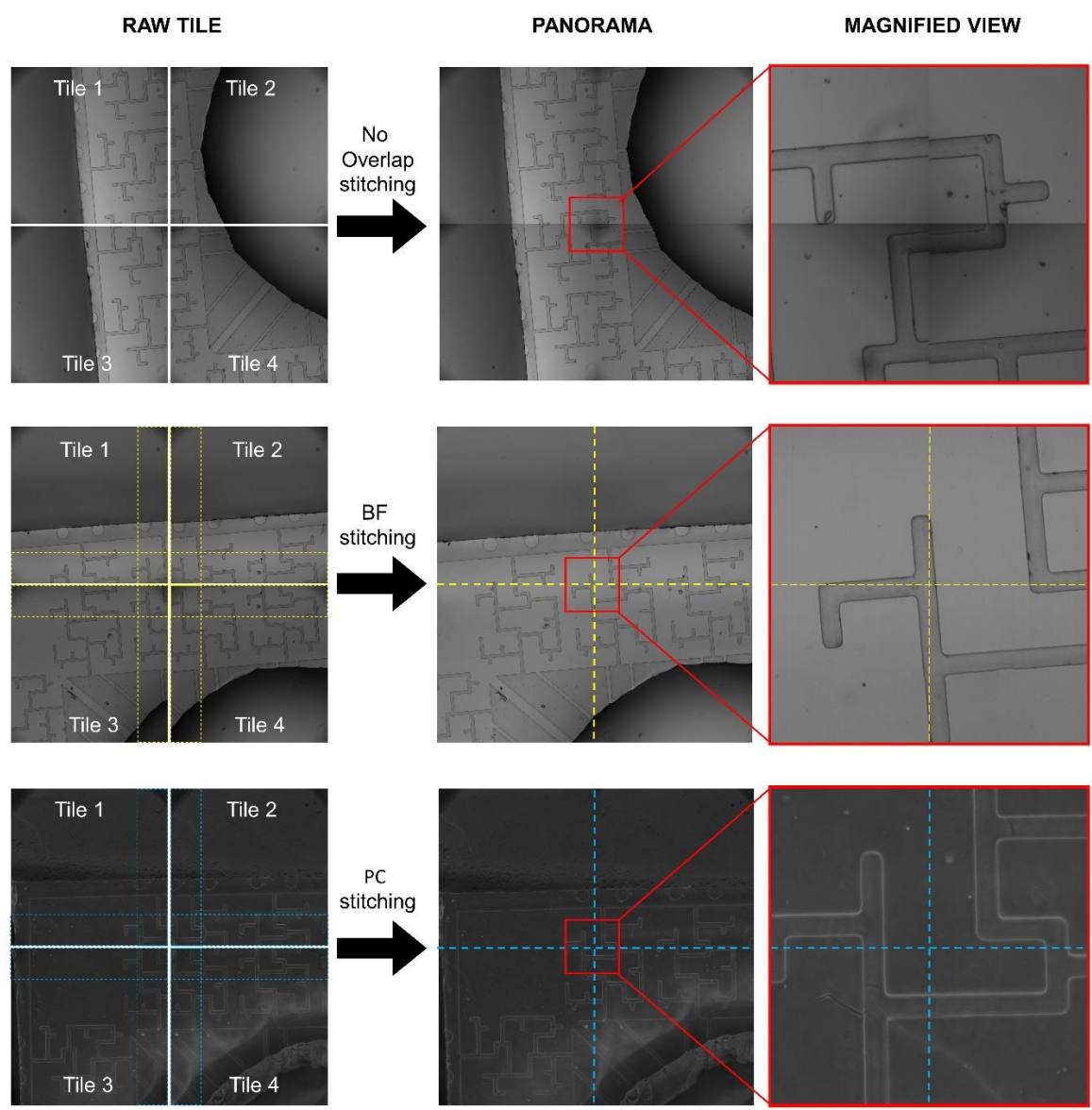

**Figure 7** Comparison between different stitching method. Top pane: Raw tiles are stitched using "No Overlap" option. Tiles are bright field images collected at 10X magnification. Middle Pane: Raw tiles are stitched using "Grid: using BF" option. Raw tiles are bright field images collected at 10X magnification. Bottom Pane: Raw tiles are stitched using "Grid: using PC" option. The tiles are generated from phase contrast mode. Dash lines in the left images show the overlaping areas between neighboring tiles. Dash line in the middle and right images indicate the border between different tiles.

In order to demonstrate the capability of the GUI to correct for shading issues, the imaging was performed with and without the "Apply Flattening" option, using bright field and phase contrast modes, acquired with 10X objective. When the flattening option is disabled, the panoramas display 2D lattice-like artifacts as seen in Figure 8A,B. These artifacts are created by the "vignette" corners of individual tiles, which is inevitable for the CMOS camera.[19] With "Apply Flattening" being enabled, the GUI eliminate artifacts using the flattening data created above. Thus, the shaded background is significantly diminished compared to the non-corrected version (Figure 8 C,D).

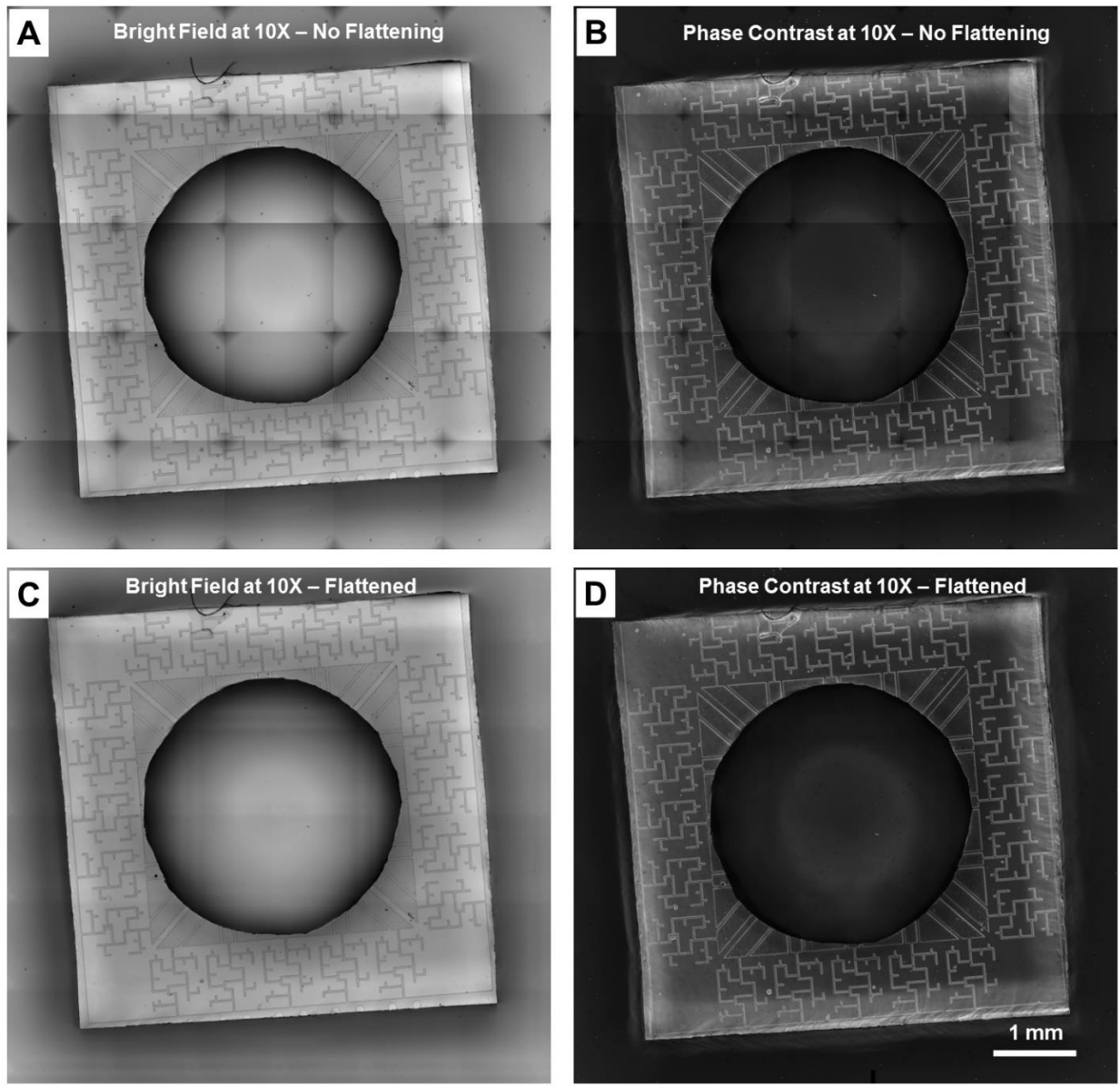

**Figure 8** Panorama produced from bright field and phase contrast image tiles obtained with 10X objective with and without flat-field correction or "flattening". (A) & (B) "Apply Flattening" option of the GUI is not selected. (C) & (D) "Apply Flattening" is enabled in GUI.

During acquisition, the mechanical movement of the stage is likely to introduce vibration which may result in blurry images and videos. Our GUI accounts for the frame-to-frame shifting of the images by using a stabilization algorithm, which compensates for the XY displacement of pixels in one image with respect to other images in the same sequence. The stabilization ability of the GUI is demonstrated in Video S1. When the "Execute Stabilization" option in the GUI is not selected, the final video appears to be "shaky" with static objects (i.e., the maze) moving from frame to frame (Video S1, LEFT). However, when the "Execute Stabilization" is enabled by the user, significant enhancement is achieved with static objects displaying negligible motion as seen in Video S1, RIGHT.

**CONCLUSIONS**

We introduced a Matlab-based custom GUI which can simultaneously communicate with the open-source, cross-platform program µManager for synchronizing hardware and with other software for image processing. It can cover a wide variety of basic controls of automated microscopes, digital cameras, motorized stages, illuminators, and other microscope accessories necessary for time-lapse imaging. These controls include automatically translating the XY stage; capturing images at a predefined frequency; switching between objective lenses, condenser optics, and filter cubes; autofocusing; and adjusting illumination. In addition, the GUI offers options for image processing, such as stitching image tiles for panorama generation, stabilizing time-lapse videos, and flat-field correction by background subtraction. Step-wise instructions for using the GUI were provided, and a demonstrative fibroblast chemotaxis experiment was described. In this experiment, images were automatically acquired on pre-defined regions over 18 hours at a 10-min interval. The resultant panoramas were stitched from image tiles attained with bright-field and phase contrast microscopies, and they showed seamless features due to the application of flat-field correction. The GUI's image processing functions, such as flat-field correction and video stabilization, significantly enhance the quality of the final images and videos. These results suggest that this GUI is useful for automating the process of generating high-quality, high-throughput time-lapse images and videos. Since this software is compatible with a wide range of hardware, it can be used by many biological research labs worldwide. Additionally, the Matlab source code is open to modification and upgradation, providing researchers with a foundation for adding in their own features, such as support for additional hardware control (i.e., microfluidic pumps, valves, and sensors) and software applications (i.e., data analysis, computational simulation, computer vision, etc.).

**SUPPLEMENTAL INFORMATION**

**Video S1.** Comparison between time-lapse images acquired without stabilization (LEFT) and with the "Execute Stabilization" option selected by the user (RIGHT). Phase contrast images are acquired at 10-min interval for 18 hours using a 10X objective.

**ACKNOWLEDGEMENT**

We acknowledge the financial support from the Gustavus and Louise Pfeiffer Research Foundation.

**CONFLICT OF INTEREST STATEMENT**

We have no conflict of interest to declare.